\documentclass[useAMS,usenatbib]{mn2e}
\usepackage[english]{babel}

\usepackage{fancyhdr}
\usepackage{graphicx}
\usepackage{times}
\usepackage{natbib}
\usepackage{amsmath}
\usepackage{amsfonts}
\usepackage{amssymb}
\usepackage{multicol}
\usepackage{layout}
\usepackage{subfigure}
\usepackage{multirow}

\renewcommand{\d}{\mathrm{d}}

\renewcommand{\gtrsim}{\ga}
\renewcommand{\lesssim}{\la}

\newcommand{\degree}{\ensuremath{^\circ}}

\def\zsun{{\rm Z_\odot}}
\def\msun{{\rm M_\odot}}

\def\Mpch{{\rm Mpc/{\it h}}}

\def\fnl{ $f_{\rm NL}$ }

\title[Non-Gaussianities and SZ]{
Hydrodynamical chemistry simulations of the SZ effect and the impacts from primordial non-Gaussianities
}
\author[F.~Pace \& U.~Maio]{
Francesco~Pace$^{1}$\thanks{E-mail: Francesco.Pace@port.ac.uk},
Umberto~Maio$^{2,3,4}$\thanks{E-mail: maio@oats.inaf.it}
\\
${}^1$ Institute for Cosmology and Gravitation, University of Portsmouth, Dennis Sciama Building, PO1 3FX, Portsmuth, England, 
Great Britain\\
${}^2$ Osservatorio Astronomico di Trieste, Via G.B. Tiepolo 11, 34143 Trieste, Italy\\
${}^3$ Max-Planck-Institut f\"ur extraterrestrische Physik, Giessenbachstra{\ss}e 1,  D-85748 Garching b. M\"unchen, Germany\\
${}^4$ Leibniz-Institut f\"ur Astrophysik (AIP), An der Sternwarte, 16, 14482 Potsdam, Germany
}

\begin{document}

\date{Received \today; accepted ?}
\pagerange{\pageref{firstpage}--\pageref{lastpage}}\pubyear{0}
\maketitle

\label{firstpage}

\begin{abstract}
The impacts of Compton scattering of hot cosmic gas with the cosmic microwave background radiation (Sunyaev-Zel'dovich effect,
SZ) are consistently quantified in Gaussian and non-Gaussian scenarios, by means of 3D numerical, N-body, hydrodynamic
simulations, including cooling, star formation, stellar evolution and metal pollution (He, C, O, Si, Fe, S, Mg, etc.) from
different stellar phases, according to proper yields for individual metal species and mass-dependent stellar lifetimes.
Light cones are built through the simulation outputs and samples of one hundred maps for the resulting temperature fluctuations 
are derived for both Gaussian and non-Gaussian primordial perturbations.
From them, we estimate the possible changes due to early non-Gaussianities on: SZ maps, probability distribution functions, 
angular power spectra and corresponding bispectra.
We find that the different growth of structures in the different cases induces significant spectral distortions only in models 
with large non-Gaussian parameters, \fnl.
In general, the overall trends are covered by the non-linear, baryonic evolution, whose feedback mechanisms tend to randomize the 
gas behaviour and homogenize its statistical features, quite independently from the background matter distribution.
Deviations due to non-Gaussianity are almost undistinguishable for \fnl$\lesssim 100$, remaining always at few-per-cent level, 
within the error bars of the Gaussian scenario.
Rather extreme models with \fnl$\sim 1000$ present more substantial deviations from the Gaussian case, overcoming baryon 
contaminations and showing discrepancies up to a factor of a few in the spectral properties.
\end{abstract}



\section{Introduction}\label{Sect:introduction}


In the current paradigm of cosmological structure formation, stars, galaxies, and clusters of galaxies develop by gravitational 
collapse in an expanding space-time \cite[e.g.][]{Gunn1972,Press1974,White1978,Peebles1993, Sheth1999,
Peacock1999,Hogg1999,Barkana2001,ColesLucchin2002,Peebles2003,Ciardi2005,Bromm2011}, growing from primordial matter perturbations 
originated during the very early phases of the Universe, during the Inflation Era \cite[][]{Starobinsky1980,Guth1981,
Linde1990}.
Such perturbations are usually assumed to follow a Gaussian distribution \cite[e.g.][and references therein]{Komatsu2010,
Komatsu2011,Hinshaw2012,Casaponsa2011,Curto2011}, because of the central limit theorem.
However, experimental constraints and theoretical arguments (\citealt{Peebles1983,Desjacques2010,LoVerde2011, DAmico2011}) have 
often questioned this assumption and supported the idea of possible deviations from pure Gaussianity.
Recent analyses by the Planck mission \cite[][]{Planck2013_XV,Planck2013_XVI,Planck2013_XXIV} suggest small levels of non-
Gaussianities, as well.
\\
Such deviations from non-Gaussianities can be parametrized by means of a perturbative expansion of the Bardeen gauge-invariant 
potential \cite[][]{Salopek1990,Komatsu2001,Verde2010,Desjacques2010}:
\begin{equation}
\label{eq:nong}
\Phi = \Phi_{\rm L} + f_{\rm NL} \left[ \Phi_{\rm L}^2 - <\Phi_{\rm L}^2> \right]\;,
\end{equation}
with $\Phi_{\rm L}$ the {\it linear} Gaussian part, and \fnl{} the dimensionless coupling constant ruling the magnitude of the 
deviations from Gaussianity\footnote{
Because $\Phi$ depends on the local value of the Gaussian field $\Phi_{\rm L}$, this kind of non-Gaussianity is named local.
}.
\\
The effects of non-Gaussianities are expected to affect objects arisen from the evolution of high-sigma matter density 
fluctuations 
\cite[e.g.][]{Grinstein1986,Koyama1999,Zaldarriaga2000,Grossi2009.1,Wagner2010,LoVerde2011,Pace2011, Scoccimarro2012}, and, 
consequently, baryonic structures, primordial haloes, and early proto-galaxies
\cite[][]{Maio2011a}.
Furthermore, non-Gaussianities can influence the cosmic star formation history
\cite[][]{Maio2011},
the observable gamma-ray-burst rate
\cite[][]{Maio2012b},
metal pollution processes
\cite[][]{Maio2012},
and the status of the inter-galactic medium at $z\sim 3$
\cite[][]{Viel2009}.
\\
Since the gas can interact with the cosmic microwave background (CMB) radiation via Compton scattering between photons 
and free electrons \cite[][]{Compton1923} during the whole cosmological evolution, the thermodynamical state of the 
Universe might induce imprints on the signal of time-integrated quantities, like the thermal Sunyaev-Zel'dovich (SZ) 
effect \cite[][]{Kompaneets1956, Kompaneets1957, Sunyaev1969, Sunyaev1970, Sunyaev1980b, Birkinshaw1999a}.
\\
In fact, when integrating along the line of sight, $\d l$, to estimate the CMB temperature distortions, variations in 
the cosmic plasma at various epochs sum up and give different contributions to the Comptonization $y$-parameter
\cite[][]{Kompaneets1956, Kompaneets1957}:
\begin{equation}
\label{eq:y}
y= \frac{k_{\rm B} \sigma_{\rm T}}{m_{\rm e} c^2}  \int n_{\rm e} T_{\rm e} ~\d l\;,
\end{equation}
where 
$k_{\rm B}=1.38\times 10^{-16}\,\rm erg/K$ is the Boltzmann constant,
$\sigma_{\rm T} = 6.65\times 10^{-25}\,\rm cm^2$ is the Thomson cross section (i.e. the low-energy limit of Compton 
scattering), 
$m_{\rm e} = 9.11 \times 10^{-28}\,\rm g$ is the electron mass, 
$c = 3\times 10^{10}\,\rm cm/s$ is the speed of light, 
$n_{\rm e}$ the electron number density, 
$T_{\rm e}$ the corresponding temperature (much larger than the CMB temperature, $T$, in the case of interest of 
ionized gas), and 
$ (n_{\rm e} \sigma_{\rm T})^{-1} $ represents the scattering mean free path.
The related spectral change in the CMB temperature depends on the frequency, $\nu$
\cite[][]{Sunyaev1969, Sunyaev1970, Birkinshaw1999a}\footnote{
  The classical results by \cite{Sunyaev1969, Sunyaev1970} are based on the non-relativistic diffusion equation by 
\cite{Kompaneets1956, Kompaneets1957}.
For a more precise approach with relativistic corrections see
\cite{Fabbri1981,Rephaeli1995a}.
}:
\begin{equation}
\label{eq:dtt}
\frac{\delta T}{T} = y \left[~x~{\rm coth}\!\!\left( \frac{x}{2} \right) -4~\right]\;,
\end{equation}
with the dimensionless parameter $x \equiv h\nu / (k_{\rm B} T)$, and $h$ Planck constant.
The spectral distortion in Eq.~(\ref{eq:dtt}) vanishes for $x\simeq 3.83$ (i.e., at $\nu \simeq 217\,\rm GHz$), 
increases for $x \gtrsim 3.83$, and decreases for $x < 3.83$.
In the Rayleigh-Jeans limit at low-frequencies the resulting CMB temperature variation is negative and equals
\begin{equation}
\label{eq:yRJ}
\frac{\delta T}{T} \sim -2 y \qquad  (x \ll 1)\;.
\end{equation}
Thus the CMB effective temperature fluctuations drops exponentially, as $\sim e^{-2 y}$, \cite[][]{Sunyaev1970}, and
determines colder "holes" in the temperature maps of microwave background radiation, associated to large structures containing
hot gas \cite[][]{Sunyaev1972}.
\\
Quantitatively, the strength of induced CMB anisotropies will obviously depend on the amount of structures formed in 
the particular cosmological model considered, and will rely on the specific hydro-, chemo- and thermodynamical history 
expected in different models.
In a large-scale cosmological context, spatially distributed enhancements or deficits in the SZ effect could trace the 
underlying structure distribution.
However, reliable modelling of hydro and chemical properties at different cosmic epochs is required to properly 
estimate thermal gas cooling and heating in different environments having different chemical compositions.
Initial empirical or semi-analytical estimates of the SZ effect suggested relatively large CMB anisotropies between
$\sim 10^{-3}$ \cite[][]{Sunyaev1980b} and $\sim 10^{-5}$ \cite[][]{Ostriker1986,Cole1988,Bond1988,Schaeffer1988}.
Further, more detailed studies, relying on first numerical simulations with Gaussian initial matter perturbations
(\fnl$=0$), 
gave more accurate corrections of the order of about $\sim 10^{-6}$
\cite[][]{Thomas1989, Scaramella1993, daSilva2000, Springel2001Erratum, Roncarelli2007, Pace2008}.
However, the impact of non-Gaussianities (\fnl$\neq 0$) on the SZ effect, by means of hydrodynamical, chemical 
simulations has not been investigated yet.
The present work is, so far, the first of this type, as previous investigations have, thus, neglected effects from stellar
evolution, chemical enrichment and consequent metal-dependent cooling and their interplay with the background cosmological
scenario.
\\
In the following, we will estimate the Compton $y$-parameter for models making different assumptions on the \fnl{} value.
We will build simulated light cones \cite[see e.g.][and references therein]{Pace2008} from redshift $z=0$ to $z\sim 7$ along 
some hundreds randomly chosen lines of sight.
We will show how the SZ signal can probe the underlying matter distribution by performing a detailed analysis of the light cones 
obtained in different non-Gaussian, N-body, hydrodynamic, chemistry simulations of large-structure formation.
In the simulated volumes, cooling, star formation, and feedback mechanisms are addressed on the base of the local thermodynamical 
properties of the collapsing gas, by consistently following its density, temperature and chemical composition, and by taking into 
account stellar evolution for both population III and population II-I stars.
The runs have been presented and described by \cite{Maio2011} and we refer the interested reader to that work for further 
details.\\
A consistent inclusion of stellar evolution properties of cosmic gas (as in the simulations considered here) is extremely
important when evaluating the thermal SZ effect, because the latter reflects the behaviour of gas temperatures and densities.
Therefore, while dark-matter-only simulations might be ideal tools for studying clustering, mass functions 
\cite[e.g.][]{Grossi2009, Wagner2010, Scoccimarro2012} and lensing statistics \cite[e.g.][and references
therein]{Pace2011,Hilbert2012}, simple analytical or semi-analytical estimates derived on top of them would likely fail.
\\
Previous studies of the effects of non-Gaussianity on SZ \cite[see][]{Roncarelli2010} were based exclusively on semi-analytic 
estimates from dark-matter-only simulations.
The novelty of the present work, instead, relies on the proper accounting for hydrodynamical and chemistry evolution
during structure formation, mostly concerning iron and silicon abundances, which can be significantly boosted at low $z$ by 
SNIa, and carbon and oxygen species, that are expelled mainly by SNII or AGB stars on short ($\sim 10^7-10^8\,\rm yr$)
time-scales.
The different delay times of these various stellar phases play a crucial role for a suitable temporal tracking of the enrichment 
episodes.
Lack of such physical scheme would lead to severe errors in estimating the gas metal content and the consequent cooling
capabilities at all epochs.
\\
The paper is structured as follows:
after presenting the simulations in Sect.~\ref{Sect:simulations}, 
and the technique to build the light cones in Sect.~\ref{Sect:cones}, 
we will discuss the main results about SZ maps, probability distribution functions and power spectra in Sect.~\ref{Sect:results}
and, finally, we will conclude in Sect.~\ref{Sect:discussion}.


\section{Simulations}\label{Sect:simulations}


\begin{table*}
\centering
\caption[Simulation set-up]{Initial parameters for the runs considered in this paper \cite[from][]{Maio2011}.}
\begin{tabular}{lccccccc}
\hline
\hline
Runs& Box side & Particle mass [$\msun/h$] for & Softening    &\fnl  & Pop III IMF       & Pop II-I IMF\\
    & [\Mpch]  & gas ~(dark matter)             & [kpc/{\it h}]&     & range [M$_\odot$] & range [M$_\odot$]\\
\hline
Run100.0    & 100  & $3.39\times 10^8\quad$ ($2.20\times 10^9$)& 7.8 & 0    & [100, 500]& [0.1, 100]\\
Run100.100  & 100  & $3.39\times 10^8\quad$ ($2.20\times 10^9$)& 7.8 & 100  & [100, 500]& [0.1, 100]\\
Run100.1000 & 100  & $3.39\times 10^8\quad$ ($2.20\times 10^9$)& 7.8 & 1000 & [100, 500]& [0.1, 100]\\
\hline
\label{tab:runs}
\end{tabular}
\begin{flushleft}
\vspace{-0.5cm}
{\small
}
\end{flushleft}
\end{table*}

We consider three simulations with different non-Gaussian parameters, described in \cite{Maio2011}, with initial conditions 
generated according to Eq.~(\ref{eq:nong}) with \fnl=0, 100, 1000.
Even if these non-null values of \fnl{} are somewhat larger than those provided from recent measurements using the Planck
satellite \cite[][]{Planck2013_XXIV}, our choice will allow us to better highlight the interplay of dark-matter non-Gaussianity
with gas and stellar physics.
We will be able to check how relevant contaminations due to baryon evolution and feedback effects are and how much they affect 
the resulting SZ signal in different \fnl{} models.
This will give us hints about the disentanglement of possible degeneracies between the luminous and the dark sectors, as well.
\\
The simulations were performed by using a modified version of the parallel tree/SPH Gadget-3 code \cite[][]{Springel2005}, 
which included gravity and hydrodynamics, with radiative gas cooling \cite[][]{Sutherland1993,Maio2007}, 
multi-phase model for star formation \cite[][]{Springel2003}, 
UV background radiation \cite[][]{Haardt1996}, 
wind feedback \cite[][]{Springel2003,Aguirre2001}
and metal pollution from Pop III and/or Pop II-I stellar generations, all ruled by a critical metallicity threshold of
$Z_{\rm crit}=10^{-4}\,\zsun$
\cite[see further details in][]{Yoshida2003, Maio2006, Maio2007, Tornatore2004, Tornatore2007, Tornatore2010, Maio2010,
Maio2011b, Petkova2012, Maio2011c, Biffi2013}.
We stress that in the simulations hydrodynamical quantities are self-consistently estimated by taking into account the proper 
yields (for He, C, O, Si, Fe, S, Mg, etc.) from stellar evolution during AGB, SNII, SNIa phases and metal-dependent (resonant 
and fine-structure) cooling rate for each particle and at each time step.
The initial mass function for PopII-I star formation regime is assumed to be Salpeter over the $\rm [0.1, 100]~M_\odot$ mass 
range, while for the PopIII regime it is a power-law with slope $-2.35$ over $\rm [100, 500]~M_\odot$ range.
\\
The cosmological parameters were fixed by assuming a concordance $\Lambda$CDM model with matter-density parameter
$\Omega_{\rm m,0}=0.3$, cosmological-density parameter $\Omega_{\rm \Lambda,0}=0.7$, baryon-density parameter 
$\Omega_{\rm b,0}=0.04$, expansion rate at the present epoch of H$_0=70\,\rm km/s/Mpc$ (i.e., normalized at 
$100\,\rm km/s/Mpc$, $h=0.7$), power spectrum normalization via mass variance within 8~\Mpch~radius sphere $\sigma_8=0.9$, and
spectral index $n=1$.
The cosmological volume was sampled as a cube of $\rm 100~Mpc/{\it h}$ side with resolution down to $\sim$~kpc scales at 
$z\sim 0$.
A schematic summary of the properties of the runs considered here is showed in Table~\ref{tab:runs}. Additional details are in 
\cite{Maio2011}.
\\
Gas densities and temperatures are extracted by the simulation snapshots and are projected along the line of sight to 
obtain three map samples for the three \fnl = 0, 100, 1000 cases, as described in the following Sect.~\ref{Sect:cones}.


\section{Light cones through the universes}\label{Sect:cones}


We build light cones by stacking the output snapshots and by following the procedure outlined in \cite{Pace2008}.
We make sure to cover completely the whole space from $z=0$ to $z\sim 7$, and discard possibly overlapping regions 
from different ``adjacent'' snapshots.
However, a simulation is only one realization of the many possible realizations in the Universe.
Thus, in order to have statistically meaningful results, it is necessary to avoid the unavoidable replication of the same 
structures when piling up the snapshots at different times for the same box.
We reach this goal by arbitrarily reshuffling particle positions and velocities via random rotations and translations 
of the box axes.
\\
As a result, for any choice of the seed of the random generator, we obtain maps showing different structure locations,
but keeping, on average, the same statistical properties \cite[][]{Thomas1989, Scaramella1993, daSilva2000, Springel2001, 
Springel2001Erratum}.\\
Then, at each redshift, $z$, we compute the value for $y$ in a given pixel of the map with coordinates ($i,j$),
$y^{ij}$, by discretizing Eq.~(\ref{eq:y}) on a two-dimensional grid with (physical) cell size $L_{\rm pix}$
\cite[according 
to e.g.][]{Thomas1989, Scaramella1993, daSilva2000, Springel2001, Springel2001Erratum, Roncarelli2007, Pace2008}:
\begin{equation}
\label{eqn:ydiscretization}
y^{ij} = \frac{k_{\rm B} \sigma_{\rm T}}{m_{\rm e} c^2}~\frac{V}{L_{\rm pix}^2}~\sum_{k} n_{{\rm e},k}^{ij}~
T_{{\rm e},k}^{ij}~w_{k}^{ij}\;,
\end{equation}
where $V$ is the volume discretization along the line of sight, $k$ is the summation index running over the particles in each 
pixel, and $ n_{e,k}^{ij}$, $T_{e,k}^{ij}$, $ w_{k}^{ij}$, the corresponding electron density, temperature and projected 
smoothing kernel.
We highlight that in the runs considered here electron fractions and temperatures are tracked on-the-fly, and can change, from 
particle to particle, at each time step, according to the corresponding local metal-dependent cooling and heating rates.
This is important, because in this way we get a precise estimation of $y$, taking into account the non-trivial backreaction of 
star formation, feedback effects and UV background on gaseous properties\footnote{
 In this respect, post-processing estimates of dark-matter only simulations might be misleading.
}.
\\
We repeat the procedure described above by choosing one hundred different random lines of sight to get three map 
samples for each \fnl{} value (and therefore we get a total of three hundred maps).
\\
We will denote these samples as:
${\cal S}_{\rm 0}$,
${\cal S}_{\rm 100}$, and 
${\cal S}_{\rm 1000}$,
referring to 
\fnl$=0$, 
\fnl$=100$, and
\fnl$=1000$, 
respectively.
\\
Each map covers a field of view of 1\degree.
\\
In the following sections we will show the main results about the SZ effect computed for the three different \fnl{} 
models, according to the procedure just described.
\begin{figure*}
 \centering
 \includegraphics[width=0.33\textwidth]{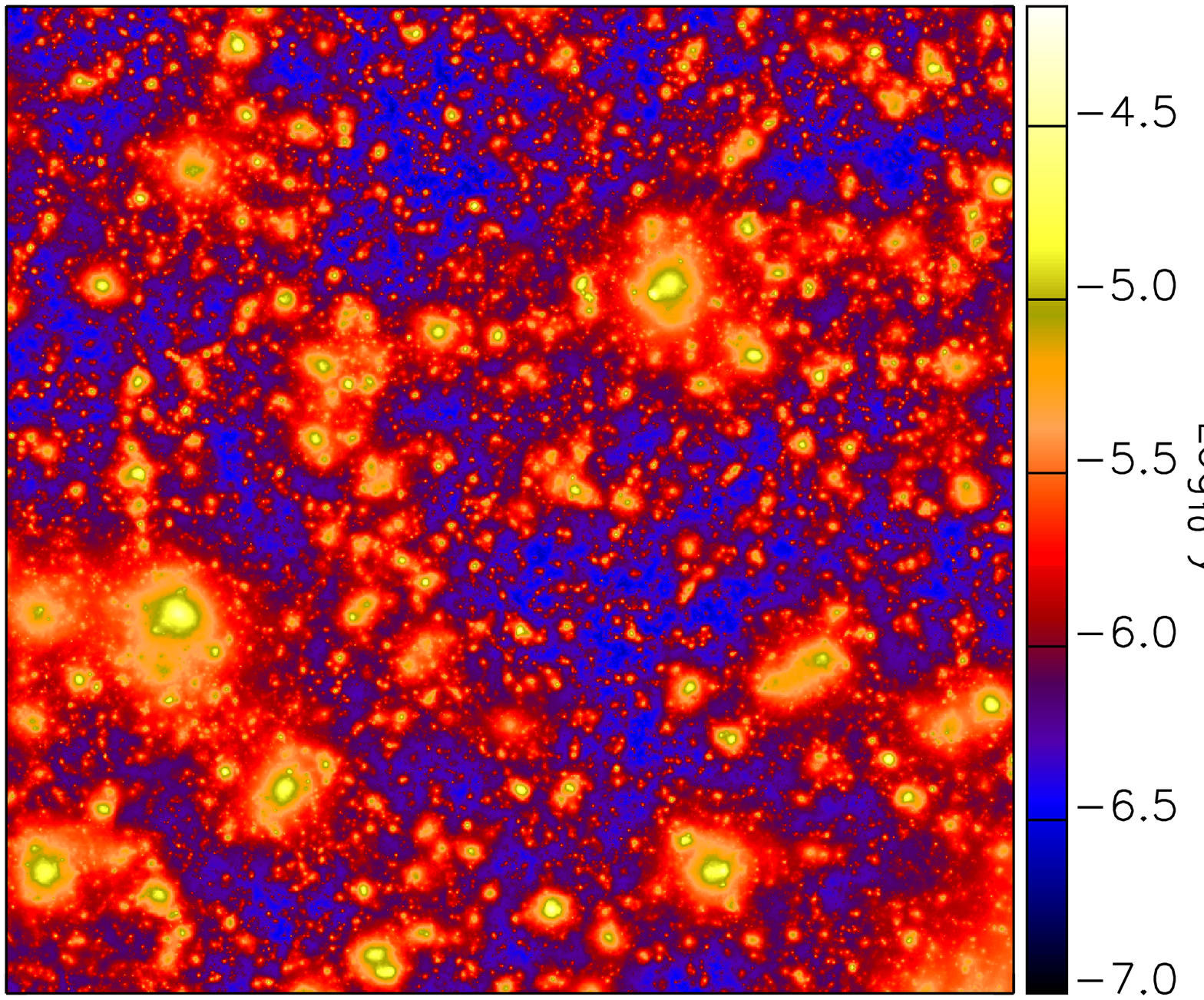}
 \includegraphics[width=0.33\textwidth]{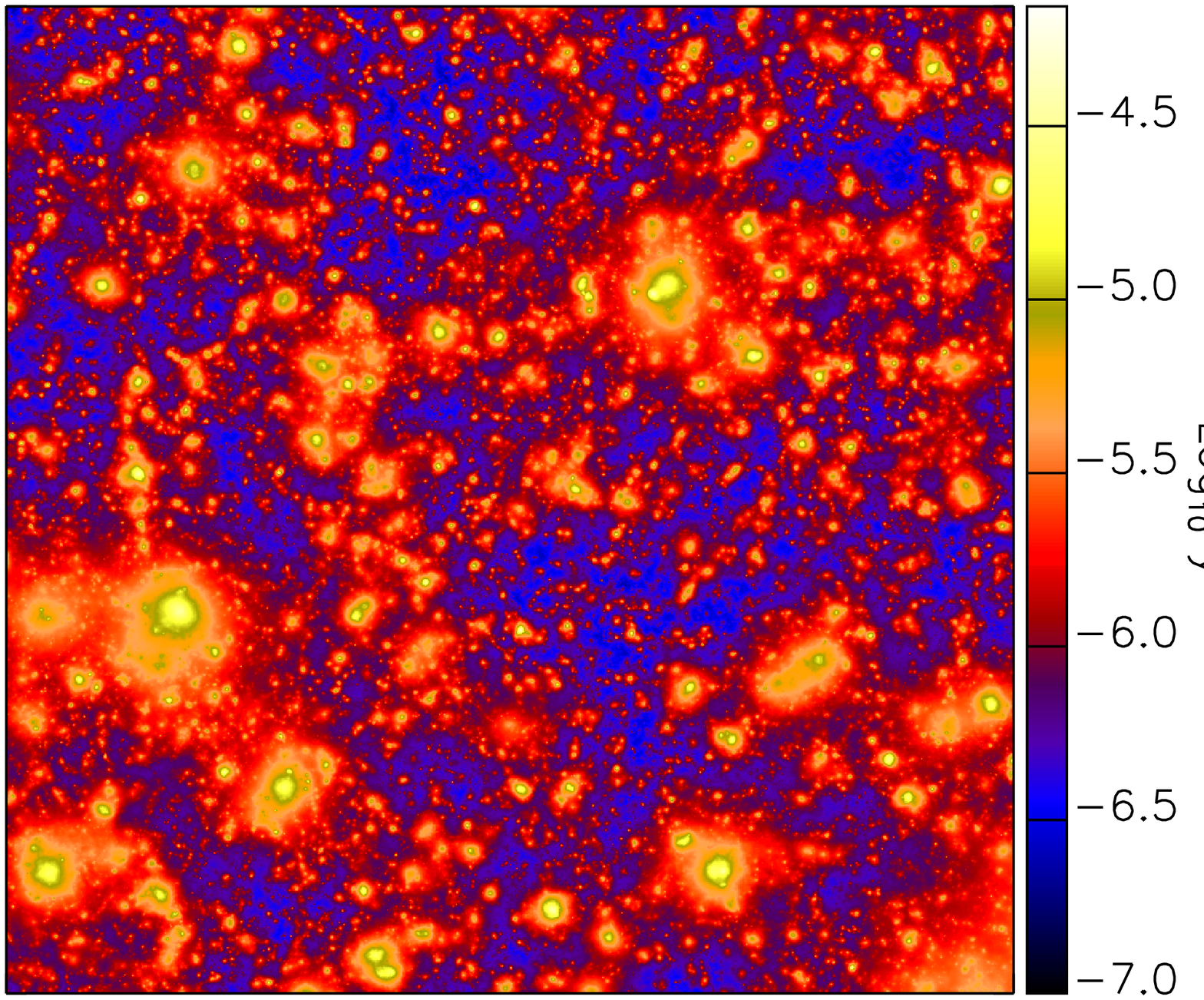}
 \includegraphics[width=0.33\textwidth]{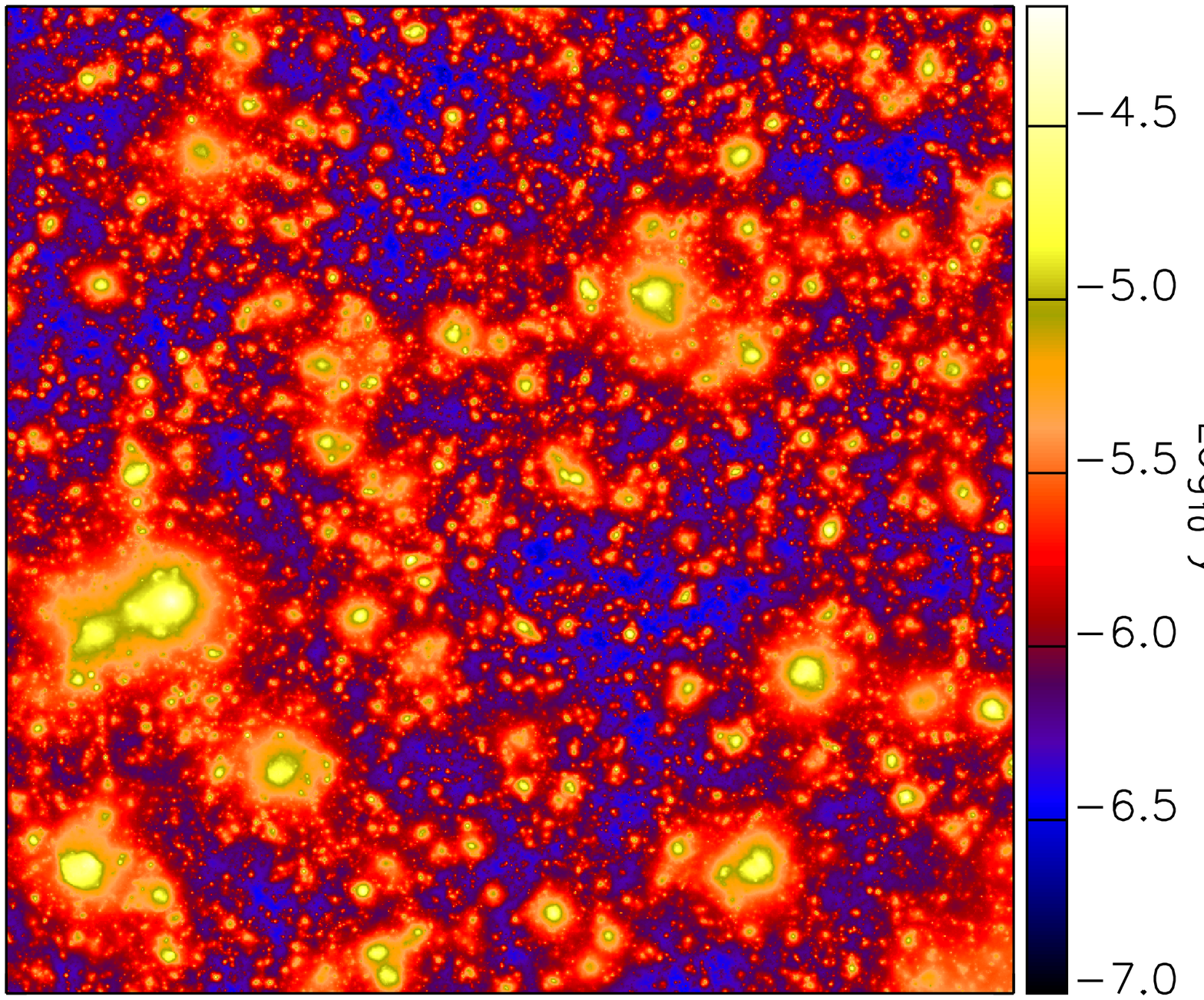}\\
 \vspace{-0.42cm}
 \fnl=0 \hspace{5cm} \fnl=100 \hspace{5cm} \fnl=1000\\
 \caption[Map comparison]{\small Comptonization $y-$~parameter maps computed, on a grid of $ 1024 \times 1024$
pixels, by integrating along the same line of sight each of the three cones obtained from the models with
    \fnl$=0$ (left),
    \fnl$=100$ (centre), and
    \fnl$=1000$ (right).
}
  \label{fig:maps}
\end{figure*}


\section{Results}\label{Sect:results}


In the following, we will first give (Sect.~\ref{Sect:maps}) a brief description of expected typical maps (both for the 
Gaussian and for the two non-Gaussian models), as obtained from the procedure outlined in Sect.~\ref{Sect:cones}, 
then we will consider the three full samples, each one made of one hundred maps and referring to the different \fnl{} 
values considered in this work.
This will allow us to analyse more carefully the statistical properties of the SZ effect and to get more solid 
conclusions about probability distributions (Sect.~\ref{Sect:szPDF}), power spectra (Sect.~\ref{Sect:szPS}) and 
bispectra (Sect.~\ref{Sect:szBS}).

\subsection{Maps}\label{Sect:maps}
In Fig.~\ref{fig:maps}, we display $y-$parameter maps for the same line of sight in the three models with
\fnl$=0$ (left),
\fnl$=100$ (center), and
\fnl$=1000$ (right), as indicated by the labels.
We note that the general structure and shape of the maps are quite similar, due to the fact that the same 
randomization process has been applied to the three cases, so the same cosmic objects broadly correspond and are easily
detectable in the three panels.
\\
While large collapsed structures are well detectable, filaments are usually not visible and covered by background signal, due 
to their typically lower densities and temperatures 
\cite[see also discussions in e.g.][]{daSilva2000, Springel2001, Springel2001Erratum, White2002c, Zhang2002, Roncarelli2007,
Pace2008}.
\\
Because of the various non-Gaussian and Gaussian initial perturbations, the growth and evolution of different structures is 
slightly different.
Typical values in Fig.~\ref{fig:maps}, are in the range between $\sim 2\times10^{-7}$ and a few times $10^{-5}$, with mean 
values of the order of $\sim 10^{-6}$.
\\
More specifically, the Gaussian \fnl$=0$ case presents a maximum $y$ of $4.70\times 10^{-5}$, while the mean is 
$1.67\times 10^{-6}$.
In the non-Gaussian \fnl$=100$ scenario, one finds a maximum of $5.13\times10^{-5}$, and the mean is $1.71\times 10^{-6}$.
In the non-Gaussian \fnl$=1000$ model, there is a maximum of $6.70\times 10^{-5}$, and the mean is $2.05\times 10^{-6}$.
This means that, while lower values for $y$ might not be significantly affected, upper values and mean values feel more 
the underlying distribution, from a few per cents up to tens-per-cent levels.
The reason for that is in the fact that lower values are found in colder environments, where the electron fraction is 
much smaller that unity and the effects of structure growth in boosting temperature and $n_{\rm e}$ are less
important.\\
More precisely, differences of $\sim 10$ per cent are found between \fnl$=1000$ and \fnl$=0$, with $y$ values in the former case 
being larger because of the more advanced heating process determined by feedback effects.
When considering the peak values, differences with the Gaussian model are evident in all the cases and reach about $\sim 8$ per 
cent for \fnl$=100$, and $\sim 30\%$ for \fnl$=1000$.
Mean values, instead, are larger than in the Gaussian case by $\sim 1.7$ per cent for \fnl$~=100$, and by $\sim 18$ per cent for
\fnl$=1000$.\\
We also mention that $y-$values do not follow a (log-)normal differential distribution, mostly because of the more extended 
high-\fnl tail \cite[as previously noted by e.g.][for the Gaussian case]{Thomas1989, Scaramella1993, daSilva2000}.
It is interesting that this conclusion still holds for the non-Gaussian cases (see next).

\subsection{Probability distributions from the whole samples}\label{Sect:szPDF}
In the following we will consider the whole samples of simulated maps -- ${\cal S}_{\rm 0}$, ${\cal S}_{\rm 100}$, 
and ${\cal S}_{\rm 1000}$ -- to draw more solid statistical constraints on the expected distribution of the
$y-$parameter in the different scenarios.
\begin{figure}
\centering
\includegraphics[width=0.3\textwidth,angle=-90]{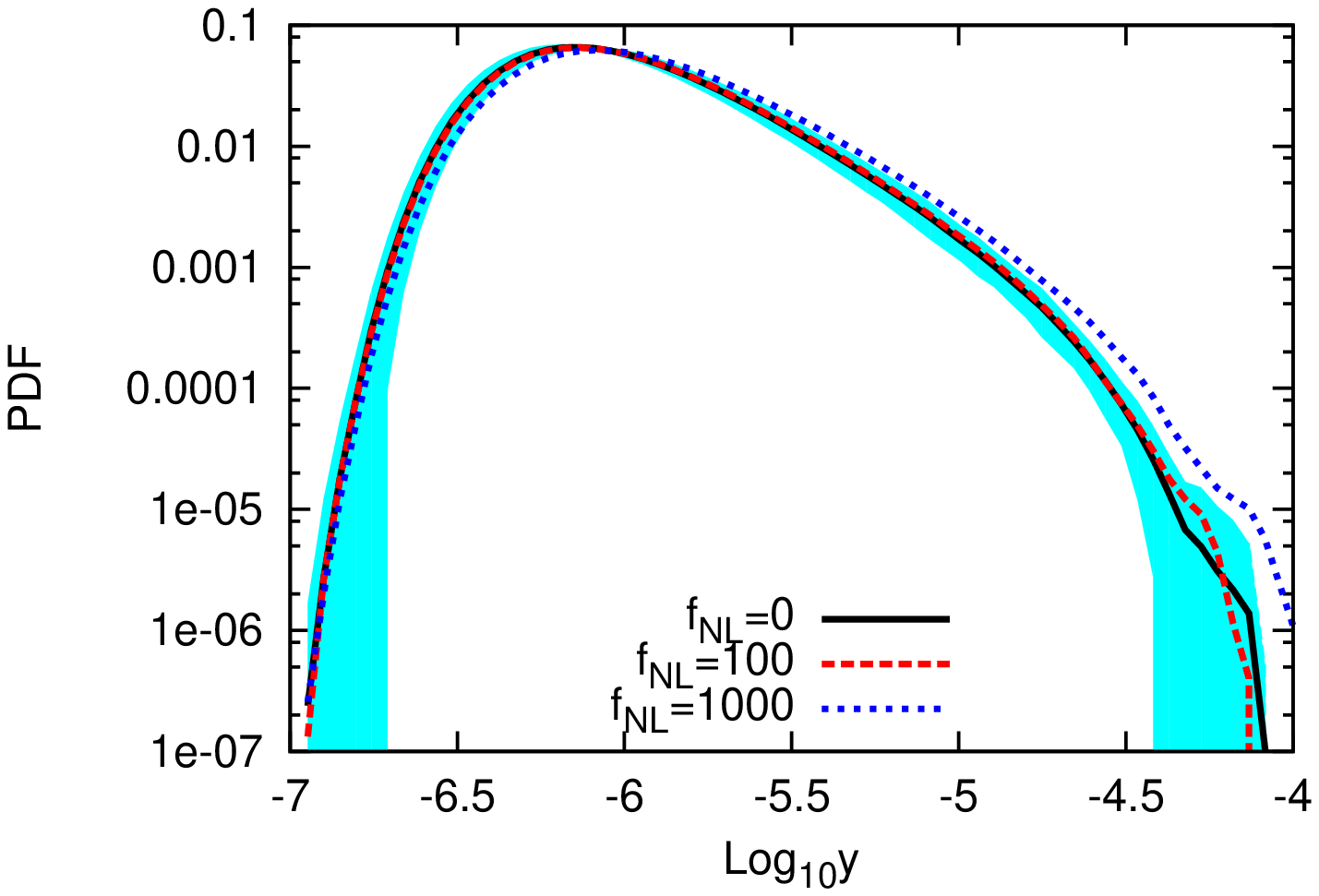}
\includegraphics[width=0.3\textwidth,angle=-90]{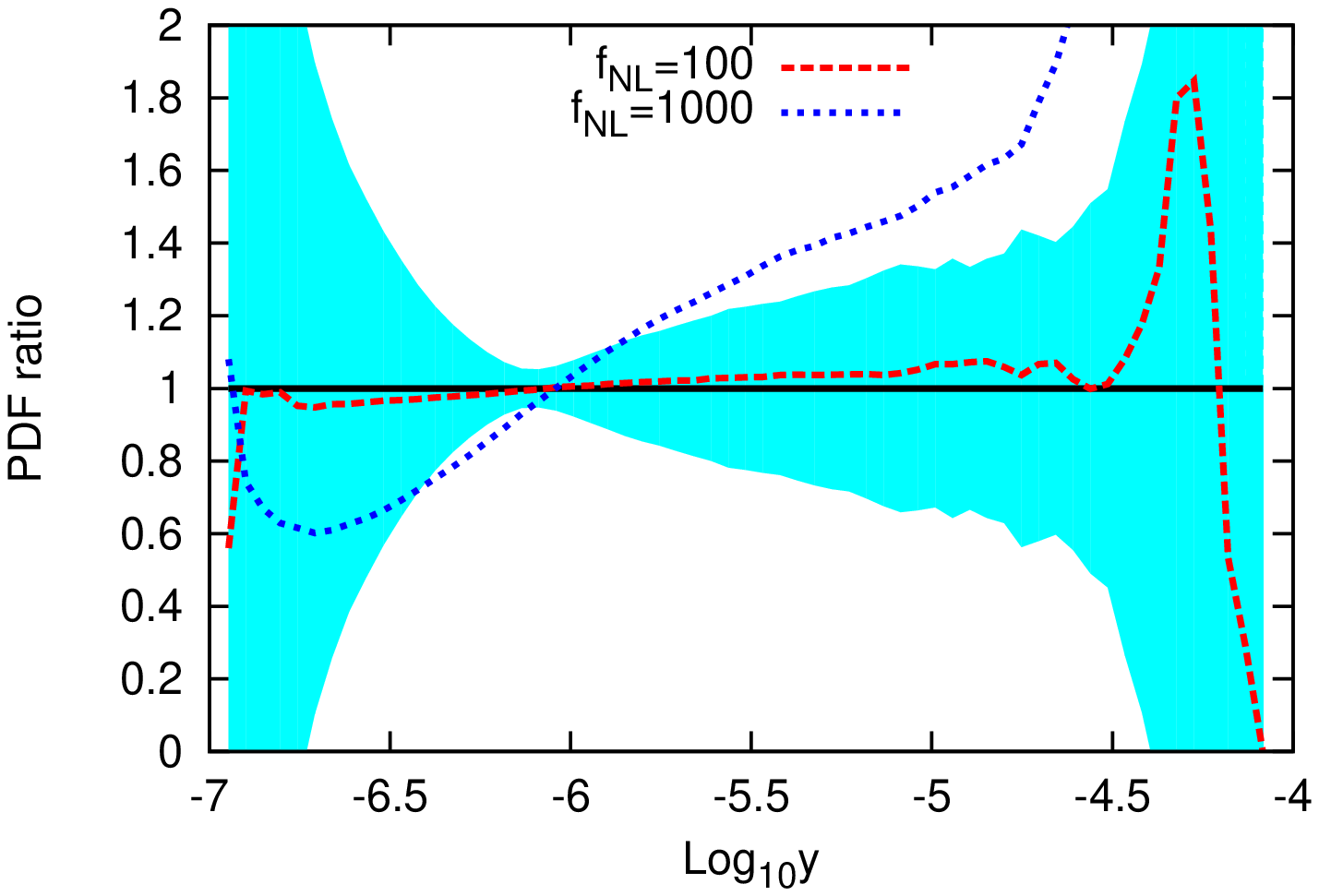}\\
\caption[PDF]{\small
Upper panel: Differential probability distribution function (PDF) for the Comptonization $y-$parameter for the three 
different models studied.
Lower panel: ratio of the PDF between the non-Gaussian and the Gaussian model.
Black line refers to the \fnl$=0$ model, red dashed line to the \fnl$=100$ model and the blue dotted line to the \fnl$=1000$
model. The shaded region has a width equal to that of the error bars of the Gaussian model.}
\label{fig:PDF}
\end{figure}
\\
In Fig.~\ref{fig:PDF} we show our results for the differential probability distribution function (PDF) for the three models as 
a result of the averaging over 100 realisations.
For sake of clarity, error bars, shown as a shaded region of equivalent width, are presented only for the Gaussian model
(black curve) and represent $1-\sigma$ deviations.
\\
Models with \fnl$=100$ and \fnl$=1000$ are shown with red dashed and blue dotted curve, respectively.
In the lower panels we show the corresponding ratio between the two non-Gaussian and the Gaussian models.
The presence of primordial non-Gaussianity is mostly important for very high and unrealistic values of \fnl.
The non-Gaussian models present peaks in the average PDF at higher values than the Gaussian one.
This is particularly evident for \fnl$=1000$ (blue-dotted curve), while the average PDF for \fnl$=100$ is only slightly shifted 
of a few per cent.
We notice that, since curves are normalised to unity, the case with \fnl$=1000$ shows a slightly lower peak: this is understood 
with the fact that this model presents higher values for the $y$-parameter -- due to the predicted more massive and hotter 
clusters -- therefore, in order to span the same area, it must have a lower peak (see previous Sect.~\ref{Sect:maps}).
This result is consistent with the average values for samples. Average values are $1.34\times 10^{-6}$ and $1.36\times 10^{-6}$ 
for the \fnl$=0$ and \fnl$=100$ model respectively, while a higher sample average of $1.6\times 10^{-6}$ for \fnl=$1000$ is 
reached.
Thus, sample averages do not differ too much and can easily be accommodated within the error bars (see next). In addition to
this, we have to take into account that measurements will suffer of uncertainties on the cosmological parameters, as well, that
will be degenerate with baryon physics.
\\
The lower panel of Fig.~\ref{fig:PDF} clearly shows that, over a scale of two orders of magnitude in the Compton parameter
($-6.5<\log_{10}(y)<-4.5$), the model with \fnl$=100$ differs of at most of $5\%$ from the Gaussian one and it is well within 
its error bars (upper panel).
This means that the two cases are basically indistinguishable.
At very low and high values of the $y$-parameter, differences become substantial, but much less significant, due to poorer 
statistics.
Larger deviations, up to a factor of a few, arise for a value of \fnl{} ten times higher, making therefore easier to distinguish 
this model from the reference one, in particular for high values of the Compton $y$-parameter.
\\
We stress that gas thermal state in the different scenarios is significantly affected by the aforementioned baryonic processes 
that take place during cosmic structure growth (star formation, stellar evolution, metal spreading, feedback effects).
These inject entropy in the surrounding medium and introduce remarkable chaotic motions in the gas, which, in turn, wash
out, 
partly (as in the \fnl$=1000$ case) or completely (as in the \fnl$=100$ case), non-Gaussian signatures and are mainly 
responsible for a similar gas evolution within corresponding cosmic structures.
\\
Our results are in good agreement with existing investigations of weak-lensing maps and effective-convergence studies probing 
the {\it total} matter distribution of collapsed objects \citep{Pace2011}.
Also for the effective convergence, underdense (overdense) regions show a ratio smaller (higher) than unity in non-Gaussian 
models with respect to the Gaussian ones, and the differences in the PDFs are comparable to the ones found here.
However, due to the impacts from baryonic processes and feedback effects that tend to homogenize gas behaviour mostly for 
\fnl$\lesssim 100$, the SZ ratios in Fig.~\ref{fig:PDF} (sensitive to gas) are slightly lower than effective-convergence ratios 
(sensitive to the total mass and thus less affected by baryons).
We also note that extreme models with \fnl$=1000$ present very strong deviations in both cases, as a result of a more clearly 
dominant contribution of the underlying dark sector over the luminous one.
\\
In the case of the Gaussian sample, we note that our average value is slightly different than the values obtained in early works
\cite[e.g.][]{daSilva2001, Springel2001}.
This is not surprising since our simulations include much more baryonic physics than previous ones.
Comparing our findings against the Gaussian model of \cite{Roncarelli2010} we note that our average values for the $y$ parameter 
are higher, despite the similar cosmology adopted.
This is due essentially to two reasons: on one side here we integrate our light-cones up to $z\simeq 7$, while
\cite{Roncarelli2010} stopped at $z\simeq 4$, on the other side differences are also partly due to the fact that here we consider
several hydrodynamical processes as cooling, star formation and especially feedback (that increases temperatures quite
rapidly) that in DM-only simulations are not included.
This highlights that high-redshift objects can still contribute to the Compton $y$ parameter when projecting along the line of 
sight.
Therefore, our results are in better agreement with the analyses by \cite{Roncarelli2007}, performed by using the hydrodynamical 
simulations by \cite{Borgani2004b}, which were integrated up to $z\approx 6$ giving $<y> = 1.19\times 10^{-6}$.
Consequently, also the location of the PDF peak in the Gaussian scenario results in good agreement with \cite{Roncarelli2007}.
\\
Direct comparisons with other analyses of the $y$-parameter in different non-Gaussian cases are not possible as there are no 
related works available in literature.

\subsection{Power spectrum from the whole samples}\label{Sect:szPS}
Given that the SZ effect contributes to the CMB power spectrum, its theoretical knowledge is of great importance.
The Compton $y$-parameter power spectrum was studied in many papers
\cite[][]{Komatsu1999, Holder1999, Molnar2000, Cooray2000, Refregier2000, Seljak2000, Springel2001, daSilva2001, Zhang2001, 
Zhang2002, Refregier2002, Seljak2002},
but never with a detailed hydro, chemical treatment for gas physics and stellar evolution in non-Gaussian scenarios.
\\
In this section we explore the effects of primordial non-Gaussianity on the expected SZ power spectrum, and we plot error bars
as for the PDF case, we use a shaded region of width identical to the error bars only for the Gaussian model, since the
non-Gaussian ones present error bars of comparable magnitude that will be omitted for sake of clarity.
Similar to what we did for the PDF, spectra are averaged over one hundred realizations and resulting
standard deviations are computed.
\begin{figure}
\centering
\includegraphics[width=0.3\textwidth,angle=-90]{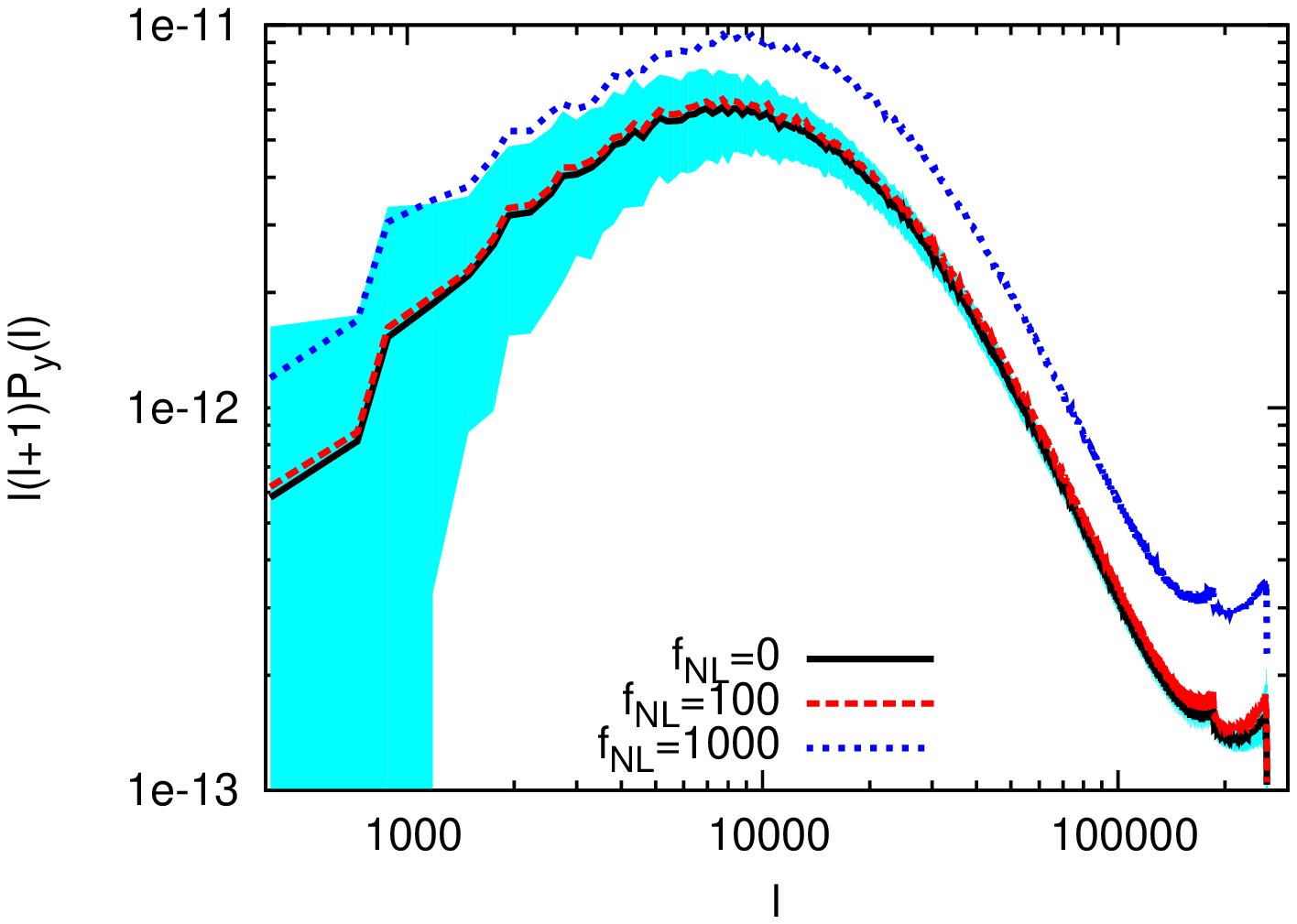}
\includegraphics[width=0.3\textwidth,angle=-90]{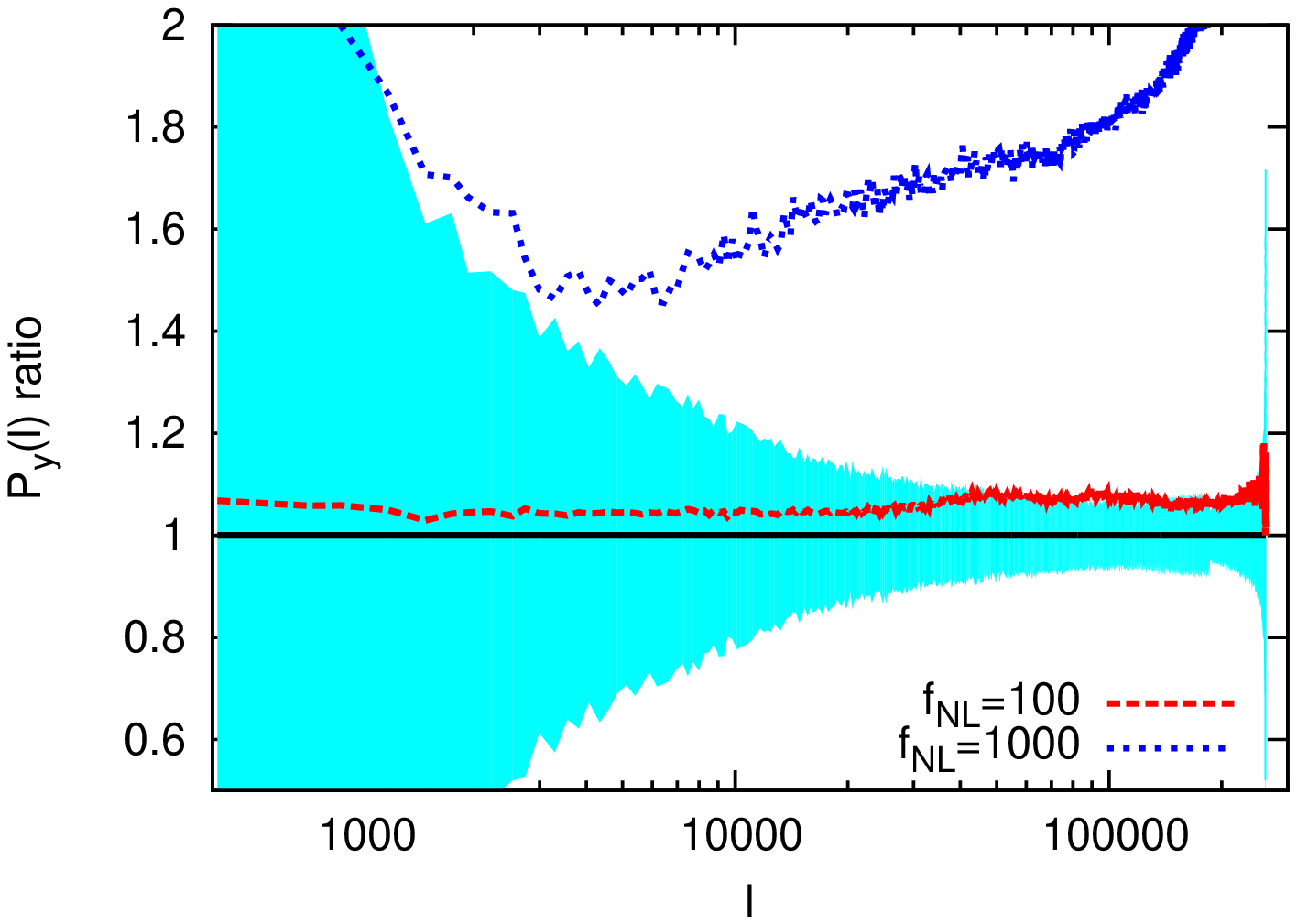}\\
\caption[PS]{\small
Upper panel: Power spectrum (PS) for the $y-$parameter for the three different models studied.
Lower panel: ratio of the PS between the non-Gaussian and the Gaussian model.
Black line with shaded region of width equal to the error bars refer to the \fnl$=0$ model, red dashed line to the
\fnl$=100$ model and the blue dotted 
line to the \fnl$=1000$ model.}
\label{fig:PS}
\end{figure}
The power spectrum represents the Fourier transform of the correlation function of $y$ between $\vec{\ell}_{1}$ and
$\vec{\ell}_{2}$ modes.
It is defined as:
\begin{equation}
 \langle\hat{y}(\vec{\ell}_1)\hat{y}(\vec{\ell}_2)\rangle=(2\pi)^2\delta_D(\vec{\ell}_{1}+\vec{\ell}_{2})P(\ell)\;,
\end{equation}
where the Dirac's delta assures that $\ell=|\vec{\ell}_1|=|\vec{\ell}_2|$.\\
The importance of the study of the SZ angular power spectrum lies in the fact that it is easier to detect than
individual clusters and it is very sensitive to the underlying cosmological properties
\cite[see e.g.][]{Komatsu1999,Seljak2002,Komatsu2002}.
Another important aspect is that it is rather insensitive to selection effects and it receives important contribution from 
outskirts regions of galaxy clusters, minimising the poor knowledge of their cores.\\
Our results are summarised in Fig.~\ref{fig:PS}, where we show average power spectra (more precisely
$\ell(\ell+1)P_{y}(\ell)$) for all the models (upper panel) and 
corresponding ratios with respect to the reference Gaussian scenario (lower panel).
As expected, higher values of primordial non-Gaussianity imply stronger deviations from the power spectrum evaluated for the 
Gaussian model, as evident from the trends in the lower panel.
\\
For the model with the highest amount of primordial non-Gaussianity we observe an increase of power from $\ell\approx 3000$, 
starting from about $50\%$ more power than the Gaussian case till a factor of two more power at the smallest scales 
($\ell\approx 2\times 10^{5}$).
This is consistent with the results on the effective convergence obtained by \cite{Pace2011} in their Fig.~{3}, as well (see 
discussion in the previous section).
The model with \fnl$=100$ differs from the Gaussian case of only $5-6$ per cent at most and at every frequency available it is 
well within the error bars (see upper panel in Fig.~\ref{fig:PS}).
We stress that the error bars of Fig.~\ref{fig:PS} are bigger for lower multipoles than for higher ones, because the number of 
possible realisations is much smaller in the former case than in the latter one.
A further comment is necessary to discuss the shape of the ratio of the power spectra.
In general, larger \fnl values present higher power, but there are some dependencies on the particular scales considered, as 
rarer bigger objects are more affected by non-Gaussianities than more common smaller ones.
For the simulation with \fnl=100 the ratio with the Gaussian calculations is approximately constant, since in the \fnl=100 and 
\fnl=0 scenarios statistical and physical effects are very similar and the resulting differences are not very pronounced.
This is not the case for the non-Gaussian cosmology with \fnl=1000, where we observe an evident U-shaped curve (this will 
happen also for the bispectrum -- see next Sect.~\ref{Sect:szBS}).
The increase of power at small scales (large $\ell$s) highlights the bias towards higher values of the initial 
perturbations in such model and the consequent higher clustering during the whole structure formation evolution.
Instead the trend for multipoles of $\ell \approx$ few thousands is the result of the non-Gaussian bias. As shown in
\cite{Grossi2009}, the halo bias in non-Gaussian cosmologies has a unique scale dependence: differences appear at large scales,
while on smaller scales the non-Gaussian bias approaches the value of the Gaussian bias. Therefore, using gas particles to trace
the underlying matter distribution, we expect to be affected by bias. This explains the declining part in the ratio between the
model with \fnl=1000 and the Gaussian case. The later increase is due to a combination of shot noise and bias due to higher
clustering.\\	
When we compare our \fnl=0 results with the (Gaussian) power spectrum by \cite{Springel2001} we note that the function 
$\ell(\ell+1)C_{\ell}$ shows a peak at $\ell \approx 8000$, in agreement with what found by those authors. 
Our findings are instead not easily comparable with the spectrum presented in \cite{Roncarelli2010}, since their highest 
frequency is $\ell=10000$ and no peak is clearly visible in those estimates.
We remind that the amplitude of the \fnl=0 spectrum in Fig.~\ref{fig:PS} is lower than the one predicted by \cite{Springel2001} 
and it is larger than the one expected by \cite{Roncarelli2010}.
As mentioned before, this is simply explained by taking into account the different gas physics included in our simulations with 
respect to the adiabatic gas of \cite{Springel2001} and the limited redshift sample of the dark-matter-only estimates by \cite{
Roncarelli2010}.
\\
AGNs \cite[][]{Roychowdhury2004,Roychowdhury2005,Scannapieco2008,Battaglia2010,Battaglia2012,Prokhorov2012} might be another 
source of contamination when distinguishing non-Gaussian models via SZ effect since mechanical feeding from AGNs can inject 
significant entropy into the Intracluster medium (ICM). Authors found that the peak of the power spectrum is shifted towards 
higher (lower) $\ell$ for lower (higher) heating times and that modifications in the power spectrum are small for $\ell\lesssim 
2000$, while they increase for higher multipoles, where a substantial reduction of the power at small angular scales was noticed. 
Moreover, the high-multipole range is very sensitive to the particular feedback recipe used. Whatever the particular prescription 
\cite[e.g.][]{Scannapieco2008,Battaglia2010,Battaglia2012,Prokhorov2012} adopted in the runs, this will be the same gas-heating 
phenomenon acting in all the various cosmological models, independently from \fnl. Hence, the consequent boost of the chaotic 
state of the IGM will increase the level of degeneracy among possible \fnl values and further erase gaseous signatures of 
primordial non-Gaussianities -- as any other feedback effect would do \cite[][]{Maio2011, Maio2011a, Maio2012}. The resulting $y$ 
distributions and spectra could suffer of systematic shifts, however, their ratio is expected to converge to the Gaussian 
behaviour more rapidly.
\\
It is worth saying that our conclusions on the ratio of the power spectra might be affected by errors on the precise cosmological 
parameters, due to the scaling of the Compton parameter and of the power spectrum with $\Omega_{\rm m}$ and $\sigma_8$, according 
to \cite[][]{Komatsu2002,Diego2004, Roncarelli2010}: $y\propto \Omega_{\rm m}\sigma_8^{3.5}$ and 
$C_{\ell}\propto\Omega_{\rm m}^2\sigma_8^7$.
This means that small uncertainties in the cosmological parameters might strongly impact the expected results and get degenerated 
with realistic values of primordial non-Gaussianity.

\subsection{Bispectrum from the whole samples}\label{Sect:szBS}
While Gaussian fields are entirely described by the PDF and the power spectrum (higher order moments are null), this is not true 
any more for non-Gaussian models, that, to be entirely characterised, would require the knowledge of all higher moments, 
corresponding to the so-called poly-spectra in the Fourier space.\\
In the following we will focus on the bispectrum, because it is related to the first non-null moment and possibly carries most of 
the physical information of non-Gaussian scenarios.
Furthermore, it is a very useful quantity to constrain cosmological parameters, especially in combination with the power
spectrum, and can help disentangle the effects of gravity from the effects of biasing 
\citep[see e.g.][]{Verde1998,Verde2000.1,Takada2004,Sefusatti2006,Sefusatti2010,Pace2011}.
\\
Compared to the power spectrum, the bispectrum depends on three frequencies such that in the Fourier space they form a 
triangular configuration.
The evaluation of the bispectrum for each single configuration is computationally expensive, therefore we limit ourselves to 
the study of the equilateral configuration, in which all the three frequencies are assumed to be the same.
\\
The bispectrum of the Compton parameter $y$ is defined as
\begin{equation}
\langle\hat{y}(\vec{\ell}_1)\hat{y}(\vec{\ell}_2)\hat{y}(\vec{\ell}_3)\rangle=(2\pi)^2\delta_D(\vec{\ell}_{123})
B(\vec{\ell}_1,\vec{\ell}_2,\theta_{12})\;.
\end{equation}
To form a triangle in the Fourier space, we require that $\vec{\ell}_1+\vec{\ell}_2+\vec{\ell}_3=\vec{0}$.
In the previous equation, $\theta_{12}$ represents the angle between $\vec{\ell}_1$ and $\vec{\ell}_2$, which, together with the 
triangle condition, fixes $\vec{\ell}_3$.
\\

\begin{figure}
\centering
\includegraphics[width=0.3\textwidth,angle=-90]{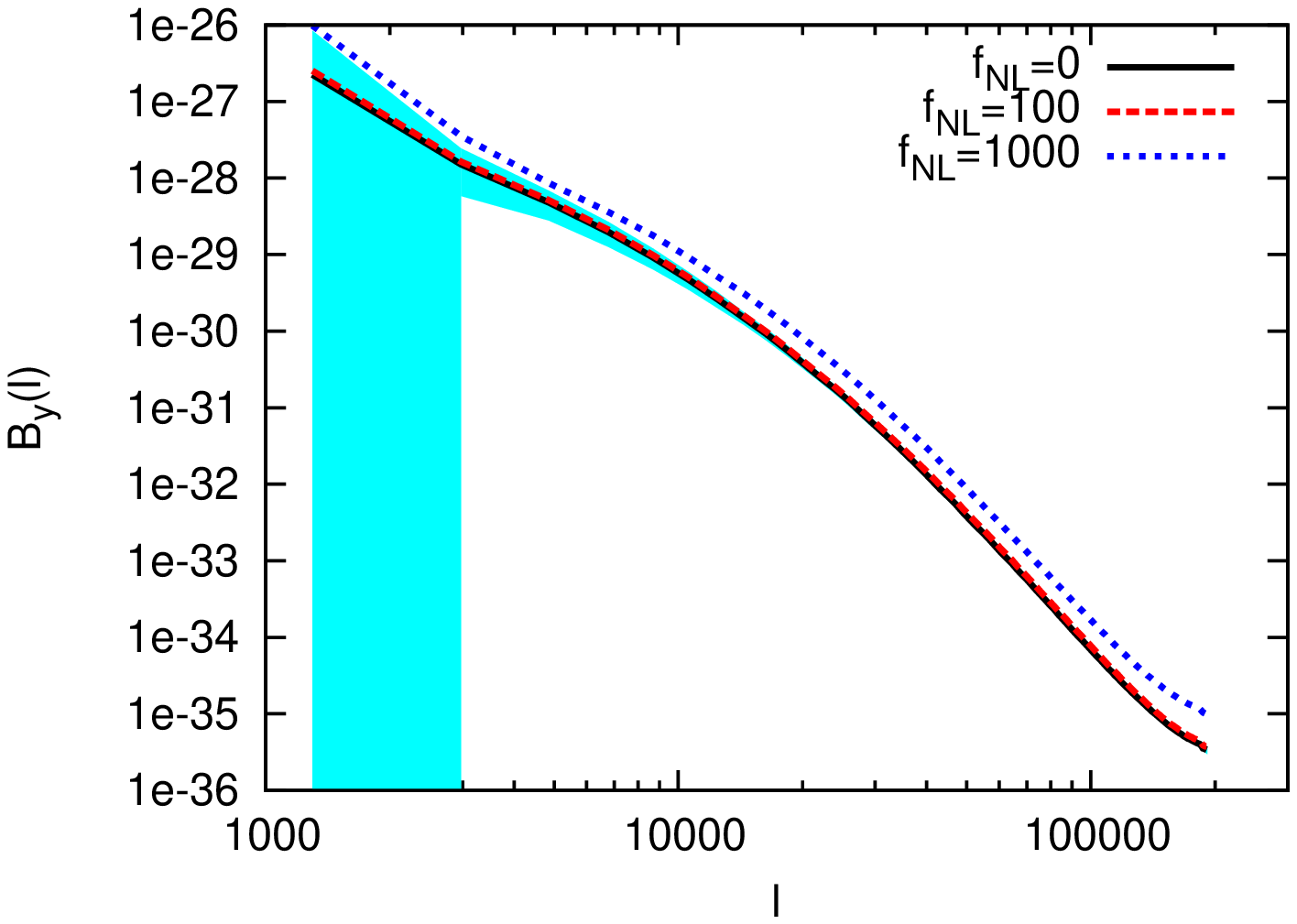}
\includegraphics[width=0.3\textwidth,angle=-90]{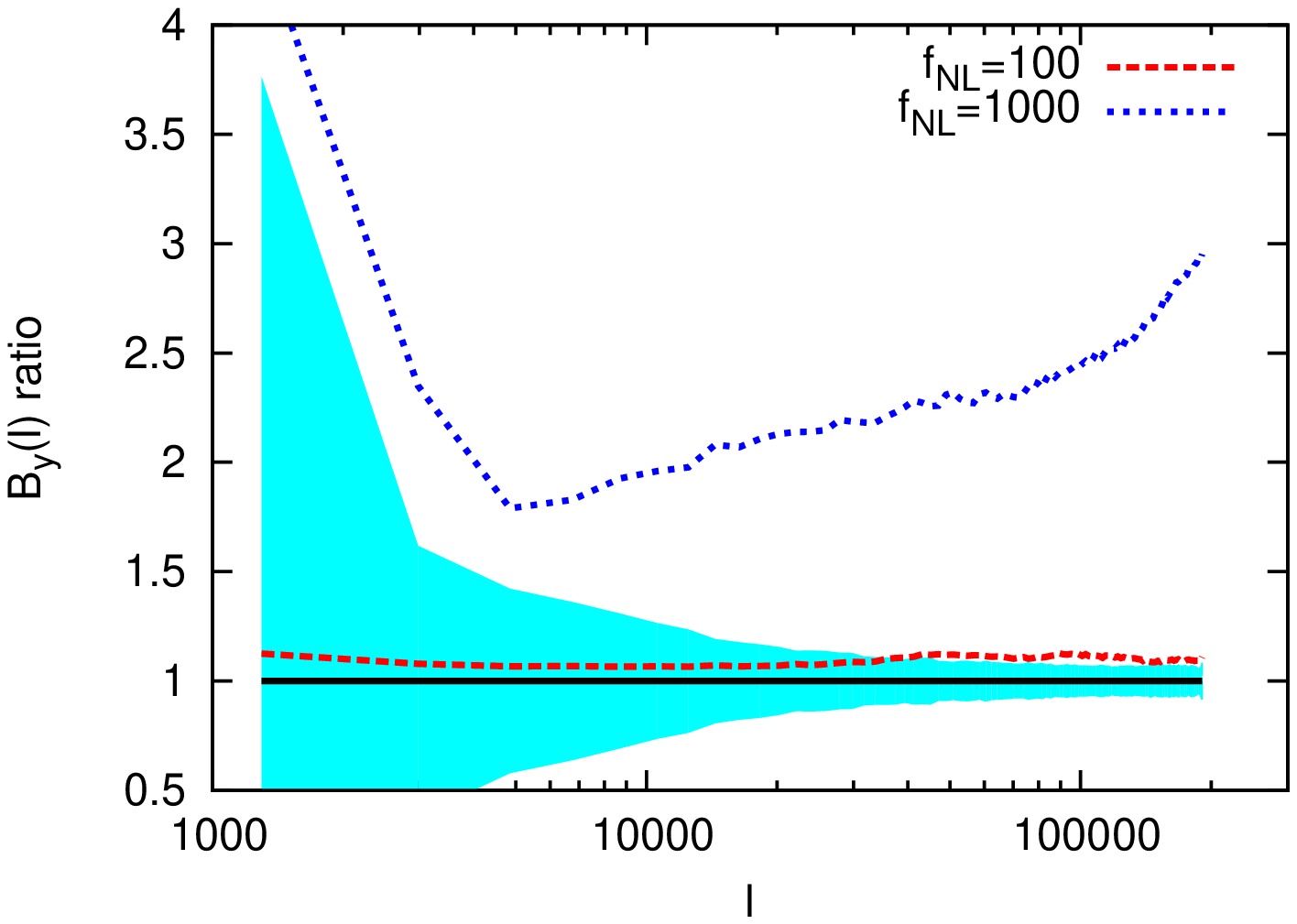}\\
\caption[BS]{\small Upper panel: Bispectrum (BS) with an equilateral configuration for the $y-$~parameter for the 
three different models studied.
Lower panel: ratio of the BS between the non-Gaussian and the Gaussian model.
Black line with shaded region of width equal to the error bars refer to the \fnl$~=0$ model, red dashed line to the
\fnl$~=100$ model and the blue dotted line to the \fnl$~=1000$ model.}
\label{fig:BS}
\end{figure}

We show our results in Fig.~\ref{fig:BS}. In the upper panel we present the comparison of the bispectra for the three different 
models studied, while in the lower panel we show the ratio between the bispectrum of the non-Gaussian models and that of the 
Gaussian model.
\\
The trends are similar to what found for the power spectrum, with relative differences increasing with the primordial
non-Gaussianity parameter.
Ratios between non-Gaussian and Gaussian bispectra are higher than the ones relative to the power spectrum since higher order 
spectra are more sensitive to deviations from Gaussianity than the power spectrum.
For the most extreme \fnl$=1000$ case the ratios at various scales range within a factor of 1.7-4, instead for the model 
with \fnl$=100$ there is a roughly constant enhancement of $\approx 10-12$ per cent, that, nevertheless, is still within the
error bars of the Gaussian \fnl$=0$ case.
\\
While for the case of the power spectrum even the model with the highest amount of non-Gaussianity considered was not so 
different from the Gaussian reference, for the bispectrum this is not the case any more.
In fact, we see that (upper panel of Fig.~\ref{fig:BS}) the bispectrum for the \fnl=1000 initial conditions is clearly off the 
error-bars at all the scales probed in our simulations.
This because the bispectrum is very sensitive to non-linearities and to clustering properties, which are enhanced in the
\fnl=1000 model.
Moreover, physically, the bispectrum is expected to scale as the square of the power spectrum and this explains values and 
shapes of the lower panel in Fig.~\ref{fig:BS}, compared to Fig.~\ref{fig:PS} (see also discussion in Sect.~\ref{Sect:szPS}).

Theoretical derivation of the bispectrum for the thermal SZ (tSZ) effect has been recently carried out by
\cite{Bhattacharya2012}. The authors use the halo model approach, as done by \cite{Komatsu2002} for the SZ power spectrum.
According to their derivation, the bispectrum is extremely sensitive to the matter power spectrum normalization
($B_{tSZ}\propto\sigma_8^{11-12}$) and to the baryon density ($B_{tSZ}\propto\Omega_{b}^4$). This has positive and negative
aspects. The positive aspect is that a combined use of the tSZ spectra (power spectrum and bispectrum) will help to reduce the
uncertainties on cosmological parameters. On the other side, such a steep dependence on the normalization is such that a small
error on $\sigma_8$ will have catastrophic consequences on the bispectrum normalisation. In other words, as shown also in
\cite{Pace2011}, the uncertainty on the cosmological parameters has by far bigger effects than primordial non-Gaussianity,
usually overcoming it.\\
Now suppose instead that all the cosmological parameters are perfectly known. The major uncertainty comes from gas physics and in
particular from AGN feedback. \cite{Bhattacharya2012} estimated a $\sim 33\%$ uncertainty on the overall amplitude of the tSZ
bispectrum, see their Fig.~5. As it looks clear from the lower panel in our Fig.~\ref{fig:BS}, errors of the order of $\sim 33\%$
in the amplitude will generically be within the error bars inferred from the different realizations up to $\ell\simeq 10^4$ and
will become progressively more important with the increase of the multipole. We also notice that therefore the uncertainty due
the gas physics will be more important than the effect of primordial non-Gaussianity, at least for \fnl=100. This shows clearly
how important is the correct inclusion of gas physics.\\
To date, the only known, at least to us, observational result on the tSZ bispectrum comes from the Planck analysis
\citep{Planck2013_XXI}. In their Fig.~11, the authors show the bispectrum for $100\lesssim\ell\lesssim 700$ for four different
configurations, equilateral, orthogonal and flat isosceles and squeezed. A direct comparison is impossible due to the different
multipoles probed here, as our bispectrum is evaluated for $\ell>1000$. Nevertheless, despite this and the very large
uncertainties, we can estimate that the amplitude of the bispectrum is comparable for both curves, making therefore our results
stronger.


\section{Discussion and conclusions}\label{Sect:discussion}


In this work we have addressed the SZ effect and the possible implications from primordial non-Gaussianities, by using suited
N-body, hydrodynamical, chemistry simulations \cite[][]{Maio2011}.
The runs include dark-matter dynamics and gas hydrodynamics, metallicity-dependent resonant and fine-structure cooling, star 
formation, feedback, stellar evolution and metal spreading according to the proper stellar yields and lifetimes.
As primordial non-Gaussianities are likely to impact the formation and evolution of dark-matter high-sigma objects and, hence, 
the whole baryonic star formation process of high-$z$ gas, induced deviations in temperatures and densities would add up when 
integrating along the line of sight and possibly show up in the behaviour of the SZ signal at $z\sim 0$.
\\
To check these issues, we build up different samples of one hundred simulated light cones, extracted from runs of structure 
formation and evolution in Gaussian \fnl$=0$ initial conditions and non-Gaussian, \fnl$=100$ and \fnl$=1000$, initial conditions.
We obtain $y$-parameter maps and study probability distribution functions, power spectra and bispectra in the different 
cosmological scenarios.
\\
In general, for mild variations from Gaussianities -- i.e. \fnl$\lesssim 100$ -- the SZ signal varies by few per cent,
while for larger variations -- \fnl$\sim 1000$ -- resulting discrepancies are much more visible and can reach a factor of a few.
\\
Minimum $y$ values are found to be not significantly affected by primordial non-Gaussianities, while mean and upper values retain 
some influence by the underlying matter distribution (see discussion in Sect.~\ref{Sect:maps} and \ref{Sect:szPDF}).
\\
These results are validated by a more general investigation of the PDF functions of the Compton parameter, $y$
(Fig.~\ref{fig:PDF}). 
The $y$ distribution for the case of \fnl$=100$ is within the error bars of the Gaussian model, instead for larger values of
\fnl$\sim 1000$ differences are more substantial.\\
We also stressed that the contribution of sources at $z>4$ is important to correctly estimate the SZ signal (see discussion in 
Sect.~\ref{Sect:szPDF}).
\\
The SZ power spectrum (Fig.~\ref{fig:PS}) in a model with \fnl$=100$ differs only of few percent from a Gaussian scenario and 
differences are within $1-\sigma$ error bars, making the two models not easily distinguishable.
Similar conclusions apply, in general, to cases with  $0<$\fnl$<100$.
In a model with ten times more primordial non-Gaussianity the underlying matter distribution and growth has a more significant 
impact on the SZ signal at all scales probed, achieving $\sim 50$ per cent or more enhancement with respect to the model with
\fnl$=0$.
Due to the detailed gas and chemical treatments, we find that, although the peak in the \fnl=0 angular power spectrum is 
consistent with early analyses \cite[e.g.][]{Springel2001}, the amplitude is lower, but in agreement with more recent estimates 
in Gaussian scenarios \cite[e.g.][]{Roncarelli2007}.
\\
The bispectrum shows a stronger signal with deviations with respect to the reference Gaussian case reaching $\sim 10-12$ per 
cent for \fnl$=100$ and even a factor of a few for the \fnl$=1000$ case (Fig.~\ref{fig:BS}).
At the same time, also error bars are bigger, and models with low \fnl{} values remain compatible with the \fnl$=0$ case.
These results are roughly consistent with the behaviour of the effective-convergence power spectra and bispectra in non-Gaussian 
models, as well \cite[see][for a deeper discussion]{Pace2011}, although weak-lensing statistics is quite insensitive to baryon 
physics and show more distinct behaviours for the different \fnl scenarios.
As shown by \cite{Bhattacharya2012}, the tSZ bispectrum is very sensitive to the matter power spectrum and its
amplitude is greatly affected by the AGN feedback. This means that effects of primordial non-Gaussianity will be overcome by the
uncertainties in the knowledge of the cosmological parameter and in the gas physics, making therefore impossible to infer
something for low values of primordial non-Gaussianity.
Thanks to the Planck satellite \citep{Planck2013_XXI} it is now possible to evaluate observationally the SZ bispectrum, but due
to the small size of our simulated box, we can not make a direct comparison since the multipoles probed in our work do not cover
the observed ones.
Despite this we notice that amplitudes of the bispectra around $\ell \simeq 1000$ are very similar.
\\
An aspect to be taken into account is the degeneracy with cosmological parameters.
Indeed, the SZ power spectrum depends both on the matter density, $\Omega_{m}$, and, much more strongly, on the matter power 
spectrum normalisation, $\sigma_8$, according to $C_{\ell}\propto\Omega_{m}^2\sigma_8^7$ \cite[e.g.][]{Komatsu2002,Diego2004}.
Thus, small uncertainties on $\sigma_8$ will affect the determination of $P(y)$ and error estimation of cosmological parameters 
could dominate the effects of intrinsic non-Gaussianity \citep{Pace2011}.
In this respect, also baryonic physics might be a severe source of contaminations, as e.g. primordial streaming motions could 
delay early star formation events and consequently alter the whole thermodynamic of collapsed objects at early times, introducing 
more degeneracies with \fnl \cite[][]{Maio2011a}.
Furthermore, stellar evolution and the final fates of stars are responsible for injecting huge amounts of entropy in the gas over 
cosmological times, directly impacting the resulting SZ signal.
As a consequence, our lack of knowledge about detailed stellar parameters, yields, initial mass function for different 
populations, feedback effects from different kind of stars, etc. might have some effects.
However, given the randomizing role of all these mechanisms, their influence should go in the direction outlined in this work, 
mostly for \fnl$\lesssim 100$ models.
\\
\noindent
In conclusion, what emerges clearly from our analyses is that implications from primordial non-Gaussianities on the SZ effect are 
strongly dependent on \fnl with larger impacts for larger \fnl values.
Scenarios in which \fnl$\lesssim 100$ are almost undistinguishable from the Gaussian counterpart.
Indeed, in these cases, the trends for the $y$ parameter PDFs, spectra and bispectra lie within the error bars of the Gaussian 
case and the discrepancies are only at a few per cent level.
More extreme models with larger \fnl values ($\sim 1000$) present more substantial deviations from the Gaussian case with 
discrepancies up to a factor of a few.

\section*{Acknowledgements}
The authors would like to thank the anonymous referee for the valuable comments that improved our manuscript.
F.~P. is supported by STFC grant ST/H002774/1. U.~M.'s research leading to these results has received funding from a Marie Curie 
fellowship of the European Union Seventh Framework Programme (FP7/2007-2013) under grant agreement n. 267251.
For the bibliographic research we made use of the NASA Astrophysics Data System.\\
Numerical computations were done on the IBM Power 6 (VIP) system at the Max Planck Computing Center Garching (RZG) and on 
the Intel SCIAMA High Performance Compute (HPC) cluster which is supported by the ICG, SEPNet and the University of Portsmouth.

\bibliographystyle{mn2e}
\bibliography{SZpaper.bbl}

\label{lastpage}
\end{document}